\documentclass[reprint,twocolumn,amsmath,amssymb,aps,superscriptaddress]{revtex4-2}

\usepackage{graphicx}
\usepackage{dcolumn}
\usepackage{bm}
\usepackage{graphicx}
\usepackage{amssymb}
\usepackage{epstopdf, epsfig}
\usepackage{yfonts}
\usepackage[colorlinks=true, linkcolor=blue, citecolor=blue, urlcolor=blue]{hyperref}
\usepackage{fontenc}
\usepackage{upgreek}
\usepackage{mathrsfs}
\usepackage{booktabs}
\usepackage{pifont}
\usepackage{amsmath}
\usepackage{bm}
\usepackage{wasysym}
\usepackage{caption}
\usepackage{subcaption}
\usepackage{euscript}
\usepackage{natbib}
\usepackage{lipsum}
\usepackage{xcolor}

\usepackage{orcidlink}

\usepackage{subcaption}
\usepackage{ragged2e}
\DeclareCaptionJustification{justified}{\justifying}
\definecolor{saulo}{RGB}{0,100,200}
\newcommand{\y}[1]{\textcolor{black}{#1}}

\newcommand{\x}[1]{\textcolor{black}{#1}}

\newcommand{\xxx}[1]{\textcolor{black}{#1}}

\newcommand{\jkx}[1]{\textcolor{black}{#1}}
\newcommand{\jkxx}[1]{\textcolor{black}{#1}}
\newcommand{\jkv}[1]{\textcolor{black}{#1}}
\newcommand{\jkvv}[1]{\textcolor{black}{#1}}
\newcommand{\vv}[1]{\textcolor{black}{#1}}

\newcommand{\ww}[1]{\textcolor{black}{#1}}
\newcommand{\vvv}[1]{\textcolor{black}{#1}}
\newcommand{\vvx}[1]{\textcolor{black}{#1}}
\newcommand{\vvk}[1]{\textcolor{black}{#1}}
\newcommand{\vva}[1]{\textcolor{black}{#1}}
\newcommand{\vvc}[1]{\textcolor{black}{#1}}
\newcommand{\vvg}[1]{\textcolor{black}{#1}}
\newcommand{\xxg}[1]{\textcolor{black}{#1}}
\newcommand{\xgg}[1]{\textcolor{black}{#1}}
\newcommand{\yg}[1]{\textcolor{black}{#1}}

\newcommand{\uu}[1]{\textcolor{black}{#1}}
\newcommand{\pp}[1]{\textcolor{black}{#1}}
\newcommand{\px}[1]{\textcolor{black}{#1}}
\newcommand{\py}[1]{\textcolor{black}{#1}}
\newcommand{\pz}[1]{\textcolor{black}{#1}}
\newcommand{\ppp}[1]{\textcolor{black}{#1}}

\newcommand{\xxk}[1]{\textcolor{black}{#1}}

\newcommand{\z}[1]{\textcolor{black}{#1}}
\newcommand{\jfm}[1]{\textcolor{blue}{#1}}

\newcommand{\sm}[1]{\textcolor{black}{#1}}

\newcommand{\jk}[1]{\textcolor{black}{#1}}

\begin{document}

\title{\jkv{Saturation} of Rogue Wave \jkvv{Amplification \vv{over} Steep Shoals}}

\author{Saulo Mendes\,\orcidlink{0000-0003-2395-781X}}
\email{saulo.dasilvamendes@unige.ch}
\affiliation{Group of Applied Physics, University of Geneva, 1205 Geneva, Switzerland}
\affiliation{Institute for Environmental Sciences, University of Geneva, 1205 Geneva, Switzerland \\ \href{https://doi.org/10.1103/PhysRevE.106.065101}{doi.org/10.1103/PhysRevE.106.065101}}

\author{Jérôme Kasparian\,\orcidlink{0000-0003-2398-3882}}
\email{jerome.kasparian@unige.ch}
\affiliation{Group of Applied Physics, University of Geneva, 1205 Geneva, Switzerland}
\affiliation{Institute for Environmental Sciences, University of Geneva, 1205 Geneva, Switzerland \\ \href{https://doi.org/10.1103/PhysRevE.106.065101}{doi.org/10.1103/PhysRevE.106.065101}}

\begin{abstract}
\uu{S}hoaling surface gravity waves \uu{induce} rogue wave formation. \uu{Though common}ly reduced to water waves passing \jkx{over} a step, non-equilibrium physics \jkvv{allows} \jkv{finite slopes \jkvv{to be} considered} \uu{in this problem}. Using non-homogeneous spectral analysis of a spatially varying \vvg{energy density ratio} we \ww{describe the dependence of the amplification as a function of the slope steepness.} \jkxx{\ww{I}ncreasing the} slope increases the amplification \jkv{of rogue wave} probability, \ww{until} \jkxx{this amplification} saturat\jkxx{es} at steep slopes. \vvg{\xxg{In contrast},  the increase of the down slope \xxg{of a subsequent} de-shoal zone lead\xxg{s to a monotonic decrease in the} rogue wave probability\xxg{, thus featuring} a strong asymmetry} \xxg{between shoaling and de-shoaling zones}. \ww{Due to \xgg{the} saturation \xgg{of the rogue wave amplification at steep slopes}, our model is applicable beyond its range of validity up to a step, thus} elucidat\ww{ing} why \ww{previous} models based on a step \ww{could describe} the physics of steep \ww{finite} slopes. \yg{We also explain why the rogue wave probability increases over a shoal while it is lower in shallower water.}
\end{abstract}


\maketitle

\section{INTRODUCTION}

\jkx{R}ogue waves \jkvv{have been observed in a variety of \vvv{fields of} physics \citep{Onorato2013}, such as astrophysics \citep{Sabry2012,Sabry2014}, optics \citep{Solli2007,Kibler2010} and condensed matter physics \citep{Efimov2010,Wen2011}.} In the ocean, they present a threat to ocean vessels and offshore operations \citep{Liu2007,Didenkulova2020}. \jkvv{In the latter case,} most studies have \jkvv{focused on} deep water, \jkv{where} \jkx{both} \citet{Benjamin-Feir1967} instability and quasi-determinism theory \citep{Boccotti2014} \jkv{apply}. \pp{The study of wave statistics evolving from deep toward shallow water regimes have become a recent trend. On the other hand, no standard distribution reproduces observations \px{over} a wide range of depths and sea states \citep{Ewans2020,Mendes2021a}. \px{F}or sea states in equilibrium (without \px{shoaling}), it may be possible to describe both deep and shallow regimes with second-order models \px{of enhanced empirical parameter space \pz{(steepness, bandwidth, depth)}} \citep{Mendes2021c,Karmpadakis2022}, \px{albeit} such methods lack first principles of the physical problem.}
\px{For seas out of equilibrium}, \jk{experiments} \jkvv{in shallower regimes have shown that inhomogeneities in the wave field due to} shoaling \jkx{contribute to} rogue wave formation \jkvv{and amplification} \citep{Trulsen2012,Raustol2014,Trulsen2020}\jkxx{.} \vvc{Paradoxically, rogue waves are less likely to occur in shallow waters as compared to deep waters \citep{Glukhovskii1966,Ewans2016}.} \pp{Furthermore, spatial statistics for seas in both equilibrium and out-of-equilibrium are not captured by available theories \citep{Mendes2020,Trulsen2022}.}

\jkxx{Recently,} \jkvv{successful} \jkxx{theories} \jkvv{of rogue wave shoaling} have arisen. For \jkxx{an} abrupt bathymetry change, \citet{Adcock2021a,Adcock2021c} \jkxx{propose} a solution in terms of the transmission coefficients and the interaction of bound waves influenced by the step. On the other hand, \citet{Moore2019} \jkv{and} \citet{Moore2020} deal\sm{t} with a step transition implementing a truncated KdV model. However, \jkx{the homogeneity of surface waves is often assumed \citep{Magnusson1996}, whereas the relaxation of this condition \jkvv{is expected to play a role in} rogue wave \jkxx{formation over a shoal} \citep{Trulsen2018}.} \jkvv{Indeed,} we \jkv{recently} provided a third framework \citep{Mendes2021b} by \jkx{taking non-}homogeneity \jkx{into account}. The\jkvv{se} \sm{three} frameworks are complementary, because \jkx{they rely on different physical approaches, respectively}: fluid dynamics analysis of wave harmonics, the statistical mechanics \jkx{of water waves}, and the \jkxx{lifting} of long-held \jkv{implicit} assumptions regarding \jkv{the} \sm{homogeneity} of ocean \jkvv{waves}. Nonetheless, the \jkxx{generality of the} third framework \jkxx{\vvc{may have} the} \sm{advantage} of being applicable to any \jkx{out-of-equilibrium} water wave problem \jkx{besides non-uniform bathymetry}\vvv{, such as opposing currents \citep{Ducrozet2021} or reflection \citep{Stuhlmeier2021}}.

\jkxx{\jkv{Although} the influence of the slope steepness on \jkv{rogue wave enhancement over a shoal} has been demonstrated in numerical simulations \citep{Gramstad2013,Adcock2020,Trulsen2021},} \jkxx{none of the three approaches \jkv{described above} consider} the \jkxx{effect of} \jkv{the} slope \ww{steepness} $\nabla h(x) \, \jkvv{\equiv} \, \partial h(x) / \partial x$ \jkxx{explicitly.} Therefore, the current work addresses analytical\vvv{ly the} problem of how the slope affects the amplification of \jkx{extreme events} when irregular waves travel \jkx{over} a shoal. We show that the slope \jkx{mainly} 
decreases the spatial energy \jkv{density} and thus increases the non-homogeneous \jkx{correction} $\Gamma$ introduced in \citet{Mendes2021b}\jkx{, increasing the rogue wave probability}. \vvv{Also,} \vv{this} slope effect \jkx{saturates} \vv{beyond a critical steepness}.
\jkx{\jkv{W}e \jkv{thus} provide a physical interpretation to} \vvv{the observation of} th\jkv{is} \jkx{saturation \jkv{by}} \citet{Adcock2020}.
\jkxx{\vv{O}ur theory explain\ww{s} why the physics of steep \ww{finite} slopes \ww{can be} well described by the three \vv{above} theories.} \ww{Furthermore, the slope effect saturates for mild slopes in shallow waters, explaining why} it is important in intermediate depths, while dying out \ww{not only} in deep \ww{water but \vvc{paradoxically} also in} shallow waters.

\section{THEORETICAL CONSIDERATIONS}

\uu{Rather than a deterministic approach \pp{based on} the hydrodynamic description of the rogue wave evolution over a shoal, the} model of \citet{Mendes2021b} \uu{uses a statistical approach focused on the integral properties of the wave system \citep{Higgins1975b}, namely the energy density. It considers the perturbation induced by the shoal on the surface elevation and \pp{thus on} the energy partition, which in turn affects the statistics of water waves.  This perturbation is spatially inhomogeneous and thereby redistributes the wave energy density throughout the bathymetry change. To derive the corresponding correction $\Gamma$ to the distribution capturing the energetic spatial evolution, we consider the} velocity potential $\Phi \jkxx{(x,z,t)}$ and surface elevation $\zeta \jkxx{(x,t)}$, \jkv{written in} generalized form \jkv{of} \jkvv{an expansion} of trigonometric functions with coefficients $(\Omega_{m \, , \, i} \, , \, \widetilde{\Omega}_{m \, , \, i})$:
\begin{eqnarray}
\nonumber
\Phi \jkxx{(x,z,t)} &=& \sum_{\jkx{m\, , \, i}} \sm{\frac{\Omega_{m \, , \, i }(k\x{_{i}} h)}{mk_{i}}} \cosh{(m\varphi)} \sin{(m \phi)} \quad ,
\\
\zeta \jkxx{(x,t)} &=& \sum_{\jkx{m\, , \, i}} \tilde{\Omega}_{m \x{\, , \, i}} (k\x{_{i}}h) \cos{(m\phi)} \quad ,
\label{eq:potentialsurface}
\end{eqnarray}
with the auxiliary variables $\varphi = k_{i} (z+h)$ and \xxx{$\phi = k_{i}(x-c_{m \, , \, i}\, t+\theta_{i})$} where \xxx{$c_{m \, , \, i}=c_{m}(k_{i})$} is the phase velocity \jk{of the $i$-th spectral component} and $m$-th order in wave steepness and $h$ the water depth. \textcolor{black}{For unidirectional waves of first-order \uu{in} steepness we extract $\Omega_{1} = a\omega / \sinh{kh}$ as well as $\widetilde{\Omega}_{1} = a$ from linear theory \citep{Airy1845}}, \jkvv{lead\uu{ing} to the energy density} \citep{Dalrymple984}: 
\begin{equation}
\mathcal{E} = \frac{1}{2} \rho g  \sm{ \sum_{i} } a_{i}^{2} \quad ,
\label{eq:energy00}
\end{equation}
where $\rho$ is the density, $g$ is \z{the gravitational acceleration}, $a$ is the \jkxx{wave} amplitude. \jkvv{A spectral energy $\mathscr{E}$ is preferred \ww{to match} the definition of power in signal processing \citep{Priestley1981,Holthuijsen2007}\vvx{, such that we define $\mathcal{E} =  \rho g \jkvv{ \mathscr{E} }$}.}
\jkx{D}ue to \jkx{the} \ww{spatial} inhomogeneity in \jkv{$\mathscr{E} $} and \jkvv{the} ensemble average $\jkv{\mathbb{E}[\zeta^{2}]}$, an initially Rayleigh \jkvv{distribution} over a flat bottom becomes $\mathcal{R}_{\alpha,\Gamma}(H>\alpha \vvx{H_{s}}) = e^{-2\alpha^{2}/\Gamma} $,
where the non-homogeneous correction \vvv{$\Gamma$ is} \citep{Mendes2021b}:
\begin{eqnarray}
\y{\Gamma (x)   = \frac{\mathbb{E}[\zeta^{2}(x , t)](x) }{ \mathscr{E}(x)   } \jkv{\approx  \frac{\langle \zeta^{\vvk{2}}(x, t) \rangle_{\vvk{t}} \vvk{(x)}  }{ \mathscr{E}(x) }  } } \quad .
\label{eq:GammaEnsemble}
\end{eqnarray}
\jkvv{Second-order waves} \jkx{have \vvg{energy density ratio}} \citep{Mendes2021b}:
\begin{equation}
\jkx{\check{\mathscr{E}}(x)} \xxg{ \equiv \frac{2\mathscr{E}(x)}{a^{2}} }  =   1+  \frac{\pi^{2}\varepsilon^{2}(x) \mathfrak{S}^{2} }{\vvk{32}}  \Big[  \Tilde{\chi}_{1}(x) + \chi_{1}(x)  \Big] 
\quad   ,
\label{eq:energydef}
\end{equation}
\ppp{where} \jkx{$\varepsilon = \vvx{H_{s}}/\vva{\lambda}$ the significant steepness of irregular waves with the significant wave height $\vvx{H_{s}}$} \ppp{(the average height of the 1/3 tallest waves)}\vvk{, zero-crossing wavelength $\vva{\lambda}$}, and
with coefficients \ww{dependent on the peak wavenumber $k_{p} = 2\pi / \lambda_{p}$}:
\begin{eqnarray}
\Tilde{\chi}_{1} =  \left[ \frac{3 - \tanh^{2}{(k_{p}h)} }{ \tanh^{3}{(k_{p}h)}  } \right]^{2}  \,\, , \,\,  \chi_{1} = \frac{9\, \textrm{cosh}(2k_{p}h) }{\textrm{sinh}^{6} (k_{p}h)} \quad .
\label{eq:energydef2}
\end{eqnarray}
\ppp{Moreover,} $\, \jkvv{1 \leqslant } \, \mathfrak{S} \, \jkvv{\leqslant 2}$ \ppp{denotes} the \ppp{slowly varying} vertical asymmetry \jkx{between crests and \jkvv{wave heights}} \ppp{($a = \mathfrak{S}H/2$)}\ppp{, which for rogue waves reads} \citep{Mendes2021a,Mendes2021b}:
\begin{eqnarray}
\ppp{ \mathfrak{S}_{(\alpha = 2)}   \approx \frac{2\eta_{s}}{1+\eta_{s}} \Bigg( 1 + \frac{ \eta_{s}  }{6}  \Bigg)   \, , \,  
\eta_{s} =  \left(   \frac{\langle \mathcal{Z}_{c} \rangle}{\langle \mathcal{Z}_{t} \rangle} \right)_{H>H_{s}}   \,  ,}
\label{eq:betalpha2}
\end{eqnarray}
\ppp{given the mean crest $\langle \mathcal{Z}_{c} \rangle$ and mean trough $\langle \mathcal{Z}_{t} \rangle$.} \jkvv{In the limit $\varepsilon \rightarrow 0$ we recover $\check{\mathscr{E}} = 1$ for linear waves.} 
\jkx{ The \ww{physics of second-order waves} \vv{leads} to} \jfm{\footnote{\jkvv{The connection between $(\Omega_{m} \, , \, \widetilde{\Omega}_{m})$ and the energies are detailed in equations 3.7-3.9 and B7-B9 of \citet{Mendes2021b}.}}}:
\begin{figure}
    \includegraphics[scale=0.53]{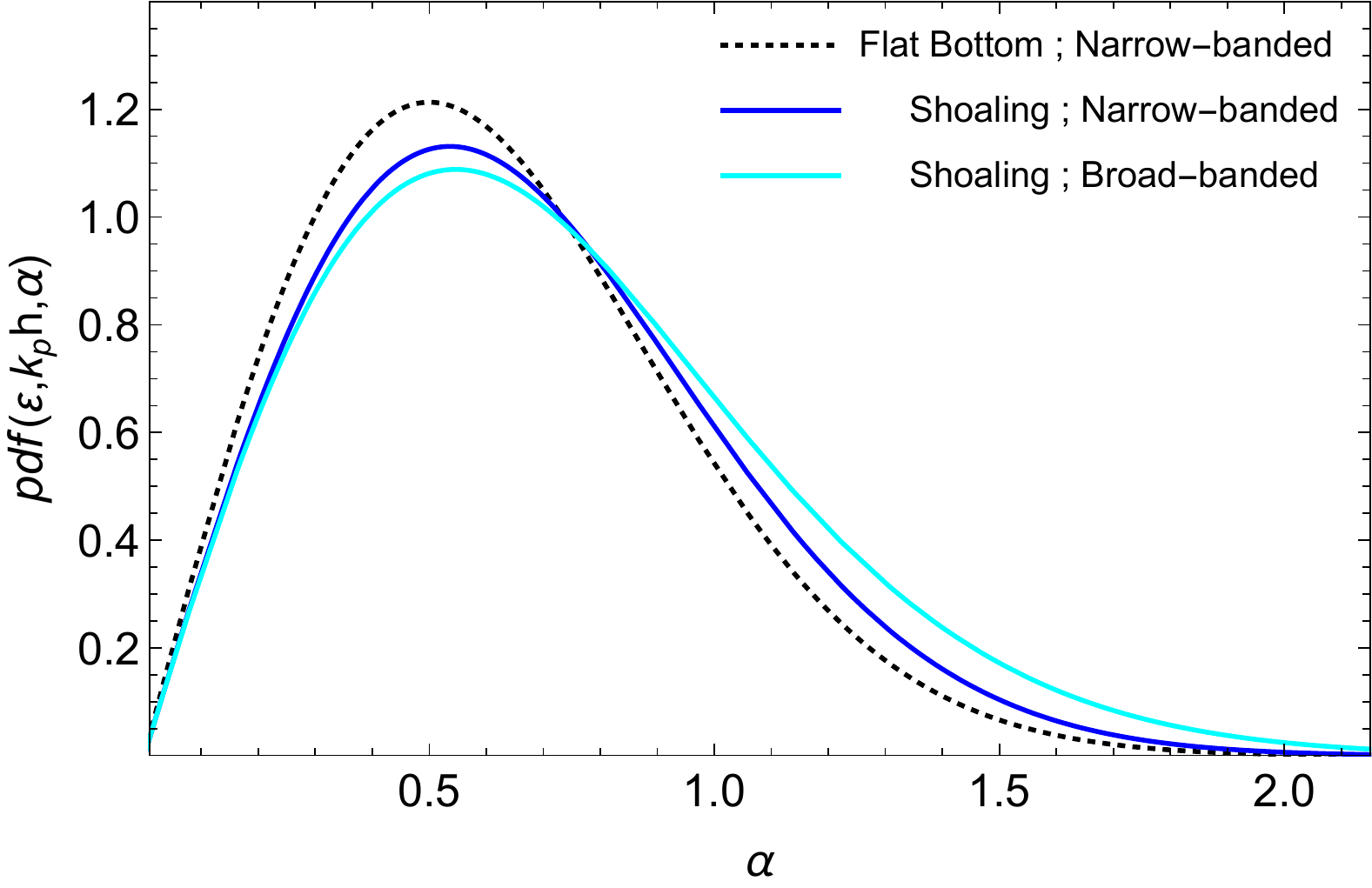}
\caption{\uu{\pp{P}robability density due to energ\py{y density} inhomogeneities \pp{caused by a shoal} as compared to the Rayleigh distribution (\pp{dotted}) for wave heights.} \pp{Shoaling featured steepness $\varepsilon = 1/20$ while broad-banded waves have $\mathfrak{S}(\alpha = 2) = 1.20$ and narrow-banded $\mathfrak{S}(\alpha = 2) = 1.05$ instead.}}
\label{fig:redist}
\end{figure}
\begin{eqnarray}
\hspace{-0.5cm}
\Gamma \equiv  \Gamma \vv{\Big(}\varepsilon (x), k_{p}h (x)\jkv{, \mathfrak{S}(x)} \vv{\Big)} = \jkv{\frac{ 1+   \frac{\pi^{2} \varepsilon^{2} \mathfrak{S}^{2}}{16}     \, \Tilde{\chi}_{1} }{  1+   \frac{\pi^{2} \varepsilon^{2} \mathfrak{S}^{2}}{32}   \, \left( \Tilde{\chi}_{1}  + \chi_{1} \right)  }} \, .
\label{eq:ee3x}
\end{eqnarray}
\uu{From the point of view \pp{of} the energy density, \px{both} a homogeneous energy density of \pp{steep} waves ($ \varepsilon \sim 1/10$) \vv{over a flat bottom} \px{and} a non-homogeneous energy density of very small waves ($ \varepsilon \ll 1/100$) \pp{over a shoal} induce a Rayleigh distribution for wave heights \citep{Higgins1952}. Otherwise, the disparity in the growth of $\mathbb{E}[\zeta^{2}(x , t)]$ and $\mathscr{E}$ will lead to a redistribution of the likelihood of wave heights (see \jfm{figure} \ref{fig:redist}), boosting the chan\pp{c}es of encountering waves \pp{with $\alpha \geqslant 0.8$} \px{and} decreas\px{ing} the chances for ordinary \pp{sized} waves \pp{$0 < \alpha < 0.8$}. Due to the vertical asymmetry $\mathfrak{S}$ \pp{between wave crests and troughs}, the repartitioning of probability \pp{is} further enhanced as the wave spectrum is broadened \pp{and waves become more nonlinear} \px{and asymmetric}. In practice, the impact of the energy repartitioning is negligible on the bulk \px{$(0 < \alpha < 0.5)$} of the exceedance probability \px{but significant for large and rogue waves ($\alpha \gtrsim 1$)}.}

\section{ANALYTICAL SLOPE EFFECT}

\begin{figure}
\includegraphics[width=8.6cm, height=5.2cm]{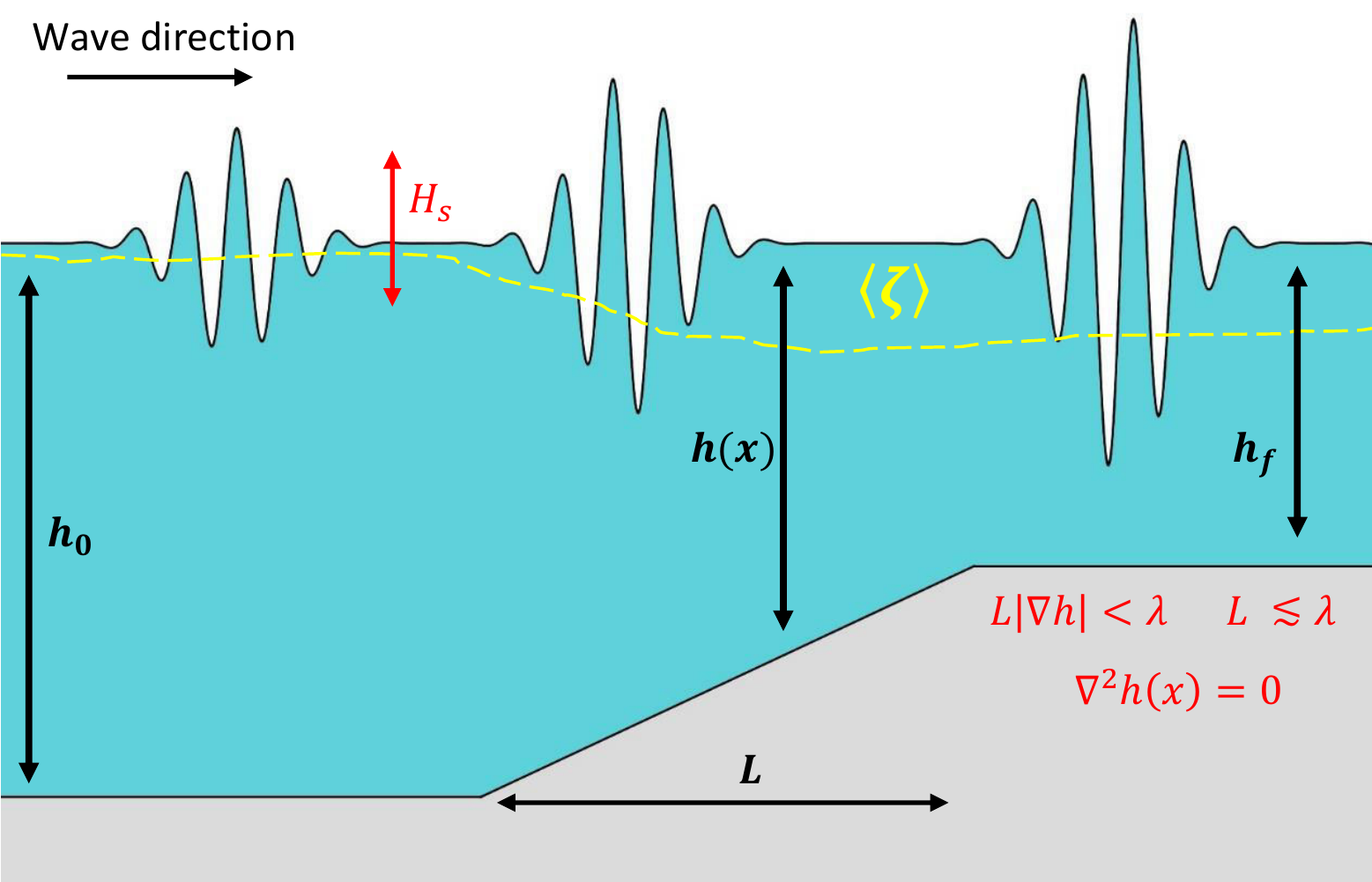}
\caption{\jkx{\ww{E}xtreme wave amplification due to a shoal \vvg{and assumptions for the solution}. \vvx{Within $x \in [0,L]$} the depth evolves as $h(x) = h_{0} + x \nabla h$ with slope $\nabla h = (h_{f} - h_{0})/L$. \yg{Note that the diagram is not to scale.}}}
\label{fig:Graphics}
\end{figure}

\uu{While our previous work focused on steep slopes, we now investigate the effect of an arbitrary finite slope.}
\jkvv{In shallow depths \vvx{($H_{s} = h_{0}$)}} \jfm{\footnote{\jkv{Surf-zone characteristics are reached when $\vvx{H_{s}} \gtrsim h_{0}$, see page 243 of \citet{Holthuijsen2007}. \vvg{\xgg{For shallow depths} we refer to a narrow \xxg{range of validity of} between Stokes and KdV realms \citep{LeMehaute1976}.}}}}, \vv{a} \jkx{constant slope} \jkvv{$\nabla h$} \jkvv{implies \vvv{an} evolution of $h(x)$} \vvx{and a\xgg{n associated} \vvg{slope-induced} set-down \xgg{($\langle \zeta \rangle < 0$)} \xgg{or} set-up \xgg{($\langle \zeta \rangle > 0$)} \xgg{effect \yg{(see} \jfm{figure} \ref{fig:Graphics}\yg{)} that affects}} the energy \uu{(and hence the rogue wave probability)} \citep{Dalrymple984} \jfm{\footnote{\vvg{While} the total energy $\vvg{\mathcal{E}}$ changes over \vvg{a} shoal \vvg{in part} because \vvg{of} changes \vvg{in} $c_{g}$\vvg{, in} the absence of wave breaking \vvc{the energy flux is constant $\nabla (\vvg{\mathcal{E}} \cdot c_{g}) = 0$}. The radiation stress will continue to increase during the shoaling \citep{Higgins1964}, thus \vv{producing} a decrease in the mean water level from $\langle \zeta \rangle = 0$ to $\langle \zeta \rangle < 0$. \xxg{If the waves break, past the breaker line we have a positive set-up $\langle \zeta \rangle > 0$.} \vvg{However}, \citet{Brevik1978} describes how other ocean processes can \vvg{induce} different types of set-down\vvg{, such as wave group or current-induced. Although the wave group set-down \xxg{over a flat bottom} is accompanied by a small return flow \citep{Higgins1962,Higgins1964}, the shoal-\xgg{induced set-down} induces a \yg{small} forward flow \citep{Higgins1983,Ehrenmark1996}, and the conservation of mass near the beach or at \xgg{the end of a} wave tank works to cancel out this forward flow \xgg{at the shoaling zone} \citep{Ehrenmark1996,Lentz2012}.}}} \uu{on top of the effect \px{previously} investigated in \citet{Mendes2021b}}:
\begin{eqnarray}
\nonumber
\vvx{ \mathscr{E}_{p}} &\vvx{+}& \vvx{\mathscr{E}_{k} }\vvx{=} \vvx{\frac{1}{2\vva{\lambda}} \int_{0}^{\vva{\lambda}} \Big\{ \Big[  \zeta (x,t) +  h(x) + \langle \zeta \rangle \Big]^{2} - h^{2}(x) \Big\} dx } 
\\
&\vvx{+}& \vvx{\frac{1}{2\vva{\lambda} g} \int_{0}^{\vva{\lambda}} \int_{-h(x)}^{\zeta} \left[ \left( \frac{\partial \Phi}{\partial x} \right)^{2} + \left( \frac{\partial \Phi}{\partial z} \right)^{2} \right] \, dz \, dx  \,\, ,}
\label{eq:energyX0}
\end{eqnarray}
\vvx{where $(\mathscr{E}_{p} , \mathscr{E}_{k})$ are the potential and kinetic energies} \jkvv{\vvx{and} we} assume \jkxx{$L \vvx{| \nabla h |} / \vva{\lambda} \vvk{\lesssim} 1$}. \jkxx{\jkvv{\vvx{The} slope \vvv{$\nabla h$} \xxg{a}ffects physical variables \vvv{such as $\nabla \lambda$}}, and the integrand of \vvx{$\mathscr{E}_{k}$} will be modified by $(x \nabla k_{p} / k_{p})^{2}$ \vvg{due to non-\xxg{negligible} derivatives from $(\partial \Phi / \partial x)^{2}$,} \jkvv{and} we find} in the limit of numerous spectral components \jfm{\footnote{\jkvv{Because $\int_{0}^{\vva{\lambda}} (x \nabla k_{p} / k_{p})^{2} dx /\vva{\lambda}$ \xgg{is} equivalent to the integral $ \int_{0}^{\vva{\lambda}} (x \nabla \vva{\lambda}/\vva{\lambda})^{2} dx/\vva{\lambda} = (\nabla \vva{\lambda})^{2}/3$.}
}}: 
\begin{equation}
\vvx{\mathscr{E}_{k}} \vvg{\xgg{\approx}} \jkx{ \sum_{m} \frac{\Omega_{m}^{2}}{4}  \cdot \frac{ \sinh{(2mkh)}}{2mgk} }  \,\,\, \vvg{,   \,\,\, \forall \,  (\nabla \lambda)^{2} \lesssim 3  \,\, .   }
\label{eq:E3}
\end{equation} 
\vvx{On the other hand, the potential energy reads:}
\begin{eqnarray}
\nonumber
\mathscr{E}_{\vvx{p}} &\vvx{\equiv}& \vvx{\mathscr{E}_{p1} + \mathscr{E}_{p2}} = \frac{1}{2\vva{\lambda}} \int_{0}^{\vva{\lambda}} \left[  \zeta^{2} (x,t) + 2 h(x) \zeta (x,t) \right] dx 
\\
&\vvx{+}& \vvx{ \frac{1}{2\vva{\lambda}} \int_{0}^{\vva{\lambda}} \left[ \langle \zeta \rangle^{2}  + 2 \langle \zeta \rangle \zeta (x,t) + 2 \langle \zeta \rangle h(x) \right] dx \quad . }
\label{eq:energy2}
\end{eqnarray}
Due to periodicity, integrals \vvk{of} $\zeta \vvk{\langle \zeta \rangle}$ \vvk{and} $ \vvk{\zeta h}$ vanish \jfm{\footnote{\vvk{The surface elevation can be decomposed into two parts $\zeta^{\ast} (x,t)= \zeta (x,t) + \zeta_{s} (x,t)$, where $\zeta (x,t)$ denotes the zero-mean oscillatory motion and $\zeta_{s} (x,t)$ the \vvg{slope-induced} set-down. However, \citet{Higgins1964} and subsequent literature did not derive \yg{the set-down $\zeta_{s}(x,t)$ explicitly}. \yg{The} periodicity of the type of solutions for $ \zeta (x,t)$ in eq.~(\ref{eq:potentialsurface}) shows that its form must be $\zeta_{s} (x,t) \sim \langle \zeta \rangle \cos^{2n}{(m\phi)}$ to fulfill $\langle \zeta^{\ast} (x,t) \rangle = \langle \zeta_{s} (x,t) \rangle = \langle \zeta \rangle$, with $n \in \mathbb{N}$. It is straightforward to show in the generalized case that $\int_{0}^{\lambda} 2 \zeta (x,t)  \zeta_{s} (x,t) dx = \int_{0}^{\lambda} 2 \zeta (x,t) \langle \zeta \rangle  dx= 0$. As long as $\nabla^{2}h = 0$ \vvg{we have $\int_{0}^{\lambda} \zeta (x,t) h(x)  dx = 0$}.
Hence, our implicit choice $\zeta^{\ast} (x,t\vvg{,n=0})= \zeta (x,t) + \langle \zeta \rangle$ \vvg{in} eq.~(\ref{eq:energyX0}) does not lead to any loss of generality.}}}. \vvg{\xgg{Moreover},} $\check{\mathscr{E}}_{p1} + \check{\mathscr{E}}_{k}$ recover\vvg{s} eq.~(\ref{eq:energydef}) \vvg{while} the slope effect on the energy is restricted to \yg{$\check{\mathscr{E}}_{p2}$}:
\begin{equation}
\vvg{ \check{\mathscr{E}}_{p2} = \frac{8}{ \mathfrak{S}^{2} h_{0}^{2} } \cdot \frac{1}{\lambda}\int_{0}^{\lambda}   \left[ \langle \zeta \rangle^{2}  + 2 \langle \zeta \rangle h(x) \right] dx  \quad ,                  }
\end{equation}
\yg{where we have used} $a = \mathfrak{S}H/2 =\mathfrak{S}H_{s}/2\sqrt{2}$.
\vvg{\xxg{Because} the set-down \xxg{is very small even} in shallow water $\xgg{|} \langle \zeta \rangle \xgg{|} / h_{0} \xgg{<} 1/20$ \citep{Bowen1968,Guza1981}, \vvg{\xxg{we find} to leading order:}}
\begin{equation}
 \check{\mathscr{E}}_{p2} \xgg{\approx} \frac{16\langle \zeta \rangle}{ \mathfrak{S}^{2} h_{0} } \int_{0}^{\lambda} \left[ 1 + \frac{ x \nabla h }{ h_{0} }  \right] \frac{dx}{\lambda}  = \frac{16}{ \mathfrak{S}^{2}  }  \left( \frac{\langle \zeta \rangle}{ h_{0}}  \right)_{\nabla h}   \left[ 1 +  \tilde{\nabla} h  \right] ,
\label{eq:primsetdown}
\end{equation}
\xgg{where \yg{$\tilde{\nabla} h \equiv \pi \nabla h / kh_{0}$ and} $f_{\nabla h}$ \yg{denotes} $f$ \yg{being} a function of $\nabla h$.} \yg{An} \xgg{ increase or decrease of the rogue wave probability is control}led by the magnitude \xgg{and sign of $\langle \zeta \rangle$} for mild slopes\textcolor{black}{, which depends on the slope} \citep{Ehrenmark1994}. \xgg{For steep slopes, the term in brackets saturates the increase in probability in the case of a shoal}.
\xxg{Following common practice, we linearize the set-down at the region near but \xgg{prior to the wave breaking zone}} \citep{Bowen1968,Guza1981} \jfm{\footnote{\xgg{Although both set-down \xgg{($\langle \zeta \rangle < 0$)} and set-up \xgg{($\langle \zeta \rangle > 0$)} depend on the slope $\nabla h$, this \yg{complex} dependence is either simplified or overlooked in the literature \citep{Saville1961,Higgins1963b,Holman1985,Gourlay2000,Hsu2006}. Indeed, the set-down/set-up are often related by a scaling of typically 1/5 \citep{Battjes1974,Gourlay2000}.}}}:
\begin{equation}
\hspace{-0.2cm}
 \nabla \langle \zeta \rangle \vvk{\approx} \frac{\nabla h}{5} \left[ 1 + \frac{8 h^{2} }{3H^{2}_{s}}   \right]^{\yg{-1}} \, \therefore \, \left( \frac{ \nabla \langle \zeta \rangle }{ \nabla h  } \right)_{ \frac{ H_{s} }{h} \ll 1  } \approx \frac{1}{5} \cdot \frac{3H^{2}_{s}}{8 h^{2} } \, .        
\label{eq:radiationstress2}
\end{equation}
\yg{Since we consider the region prior to wave breaking, the associated set-up does not develop. However, the de-shoal induces another form of set-up of smaller magnitude commonly called piling up \citep{Higgins1967,Diskin1970}.}
\vvg{Integrating eq.~(\ref{eq:radiationstress2}) over a wavelength and plugging into eq.~(\ref{eq:primsetdown}), we find \textcolor{black}{for broad-banded waves}:}
\begin{equation}
\yg{ \check{\mathscr{E}}_{p2} \approx \frac{96}{55 \mathfrak{S}^{2}} \frac{\pi \nabla h  }{ k_{p}h_{0}} \left[ 1 + \frac{\pi \nabla h  }{ k_{p}h_{0}}   \right] \approx \frac{6}{5} \, \tilde{\nabla} h \Big(1 + \tilde{\nabla} h\Big) \, .
}
\label{eq:Ep2}
\end{equation}
\xxg{However,} the \xgg{effect of depth change on} the energy \xxg{density} ratio is expected to vanish in deep water\xxg{, as} the exchange of momentum encoded in the radiation stress \ww{becom\xxg{es} negligible} \citep{Holthuijsen2007}. \xxg{Therefore}, we seek a parameterization that generalizes the slope effect to intermediate depths\xxg{, and} the energy \vvc{ratio} \vva{shall evolve} towards intermediate \xgg{depths} \vva{in} the same \vva{way as} \vvx{eq.~(\ref{eq:radiationstress2}). \xxg{T}hat is to say, $\check{\mathscr{E}}_{p2} (H_{s} \ll h_{0}) / \check{\mathscr{E}}_{p2} (H_{s} = h_{0})$ is identical to $ (  \nabla \langle \zeta \rangle / \nabla h  )_{  H_{s}  \ll h } / (  \nabla \langle \zeta \rangle / \nabla h  )_{  H_{s}  = h}$.} \xxg{We} multiply both numerator and denominator \xxg{of eq.~(\ref{eq:radiationstress2})} by $k_{p}^{2}$ \xxg{and convert the numerator} peak \xxg{wavelength} to zero-crossing wavelength \yg{$k_{p}H_{s} \approx (\pi / \mathfrak{S}\sqrt{2}) \varepsilon$} \citep{Mendes2021b} \pz{(see \jfm{figure} \ref{fig:Ep2})}:
\begin{figure}
    \includegraphics[scale=0.57]{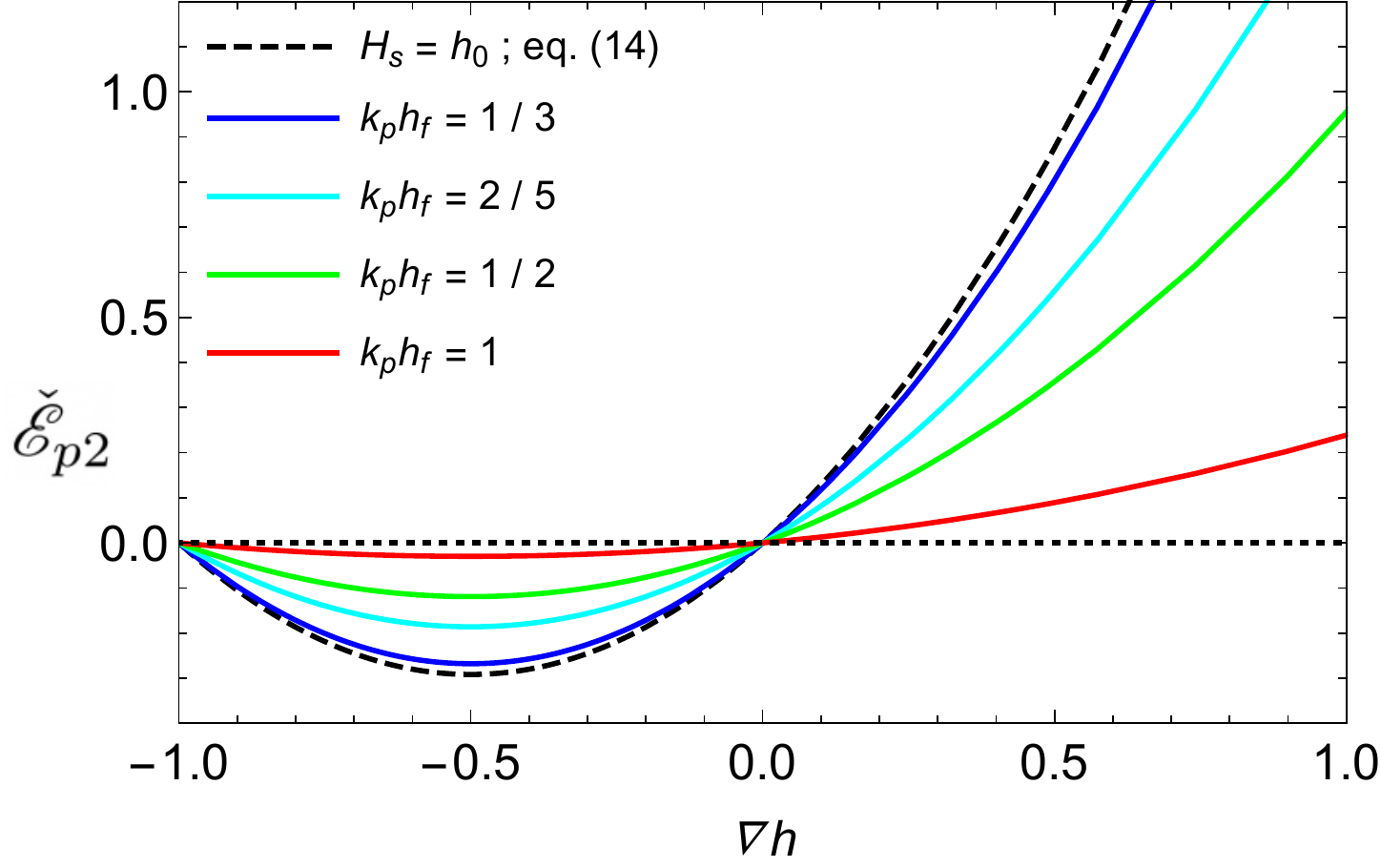}
\caption{\uu{Computation of $\check{\mathscr{E}}_{p2}$ from eqs.~(\ref{eq:energycorrec22}) (solid) and eq.~(\ref{eq:Ep2}) \py{(dashed)} for an initial depth $k_{p}h_{0}=\pi$} \pp{and $\varepsilon = 1/7$}.}
\label{fig:Ep2}
\end{figure}
\begin{figure*}[t]
\minipage{0.3\textwidth}
  \includegraphics[scale=0.43]{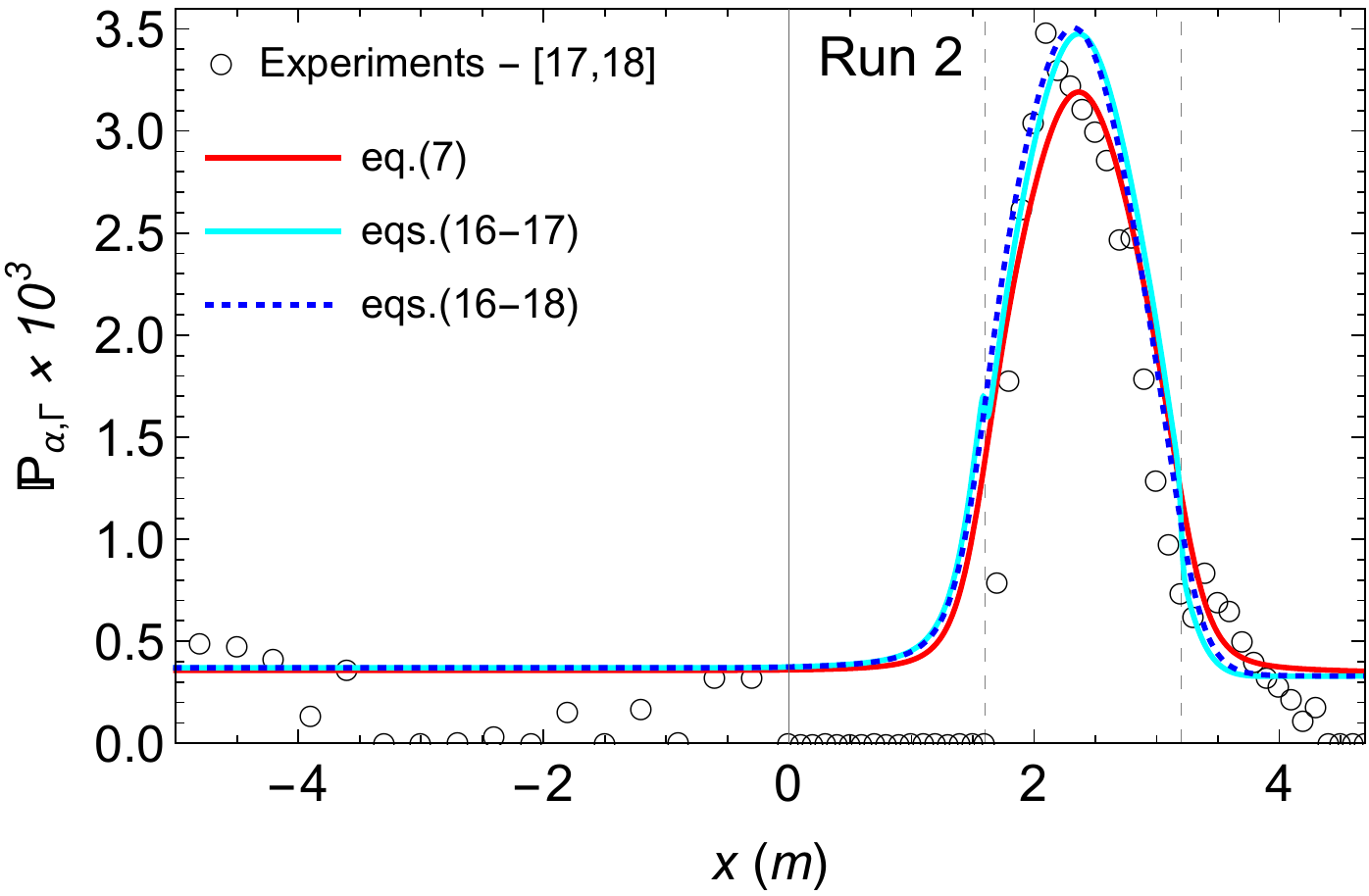}
\endminipage
\hfill
\hspace{0.0cm}
\minipage{0.3\textwidth}
  \includegraphics[scale=0.43]{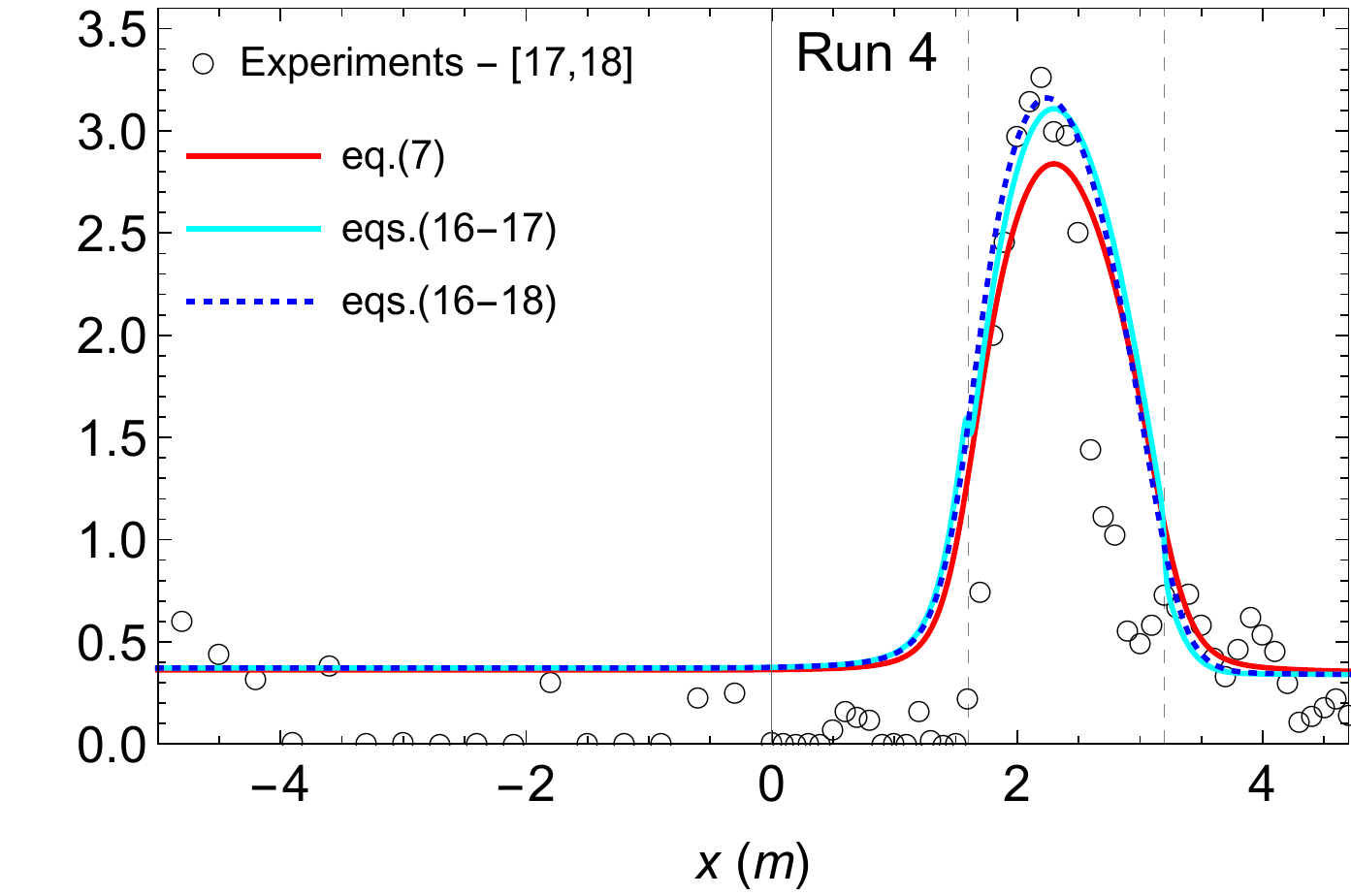}
\endminipage
\hfill
\minipage{0.35\textwidth}%
  \includegraphics[scale=0.43]{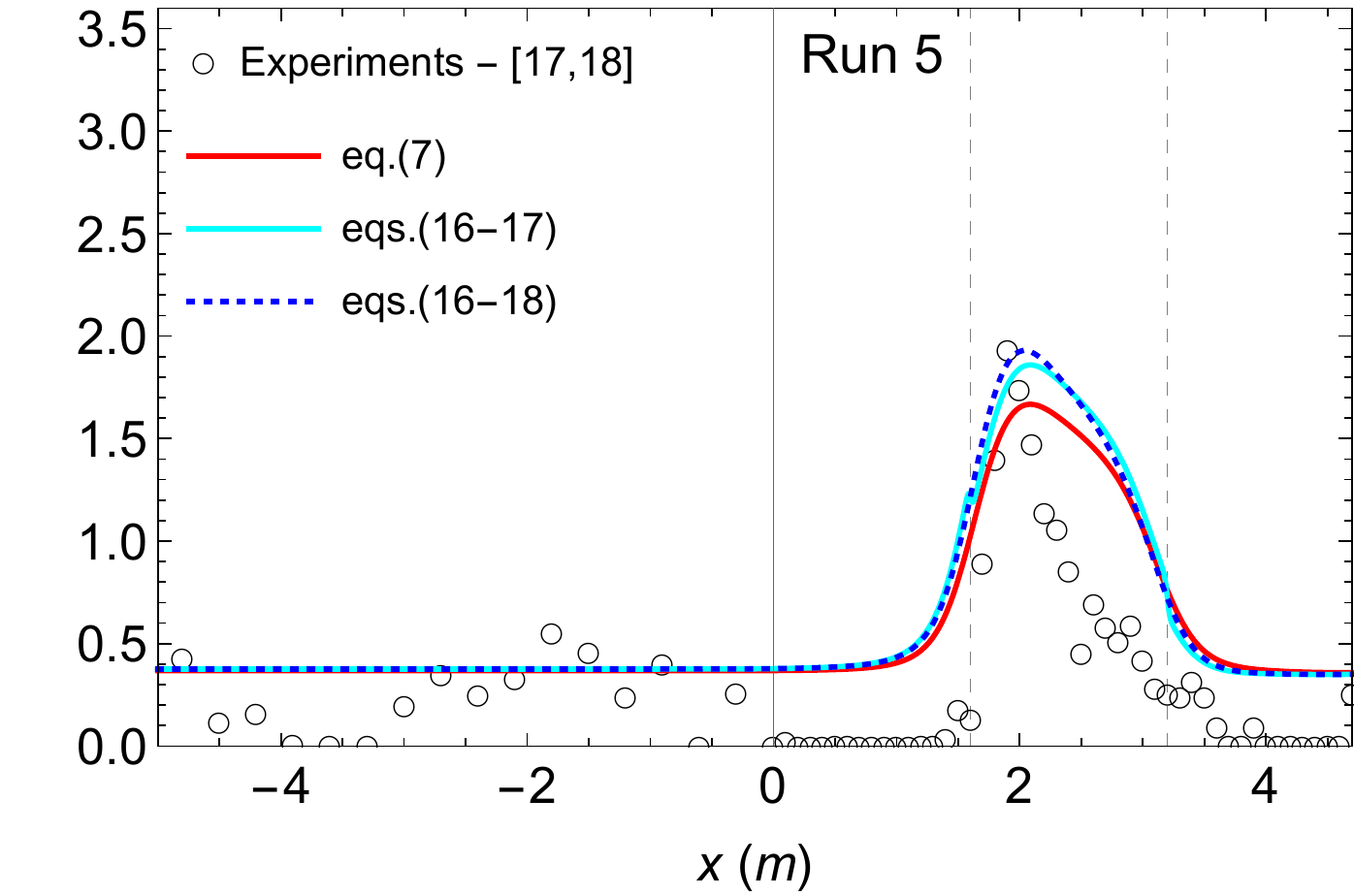}
\endminipage
\caption{Exceedance probability evolution over a bar from eq.~(\ref{eq:bath}) versus data (hollow circles) \citep{Raustol2014}. \jkv{The probability evolution has been computed} from eq.~(\ref{eq:ee3x}) \jkv{\jkvv{as in} \citet{Mendes2021b} }(solid red) and the \jkv{slope-dependent counterpart} in eq.~(\ref{eq:ee3}) without (cyan) and with (dotted blue) smoothing of the bar geometry ($\vartheta = 3$) from eq.~(\ref{eq:dh(x)}). \xxg{Note that} \jkv{the} experiments \jkv{of} \citet{Trulsen2020} \xgg{lie within} \xxg{all assumptions leading to eq.~(\ref{eq:ee3}).}}
\label{fig:TrulsenH}
\end{figure*}
\begin{equation}
\check{\mathscr{E}}_{p2}\jkv{(k_{p}h)} \approx \frac{\pi \nabla h}{k_{p}h_{0}}  \left[ 1 + \xxg{\frac{\pi \nabla h  }{ k_{p}h_{0}}}   \right] \ww{\times} \xxg{ \yg{ \frac{6\pi^{2}}{5\mathfrak{S}^{4}} } \frac{\varepsilon^{2}}{(k_{p}h)^{2}} }  \quad .
\label{eq:energycorrec22}
\end{equation}
\jkx{Plugging} eq.~(\ref{eq:energycorrec22}) \vvx{and $\check{\mathscr{E}}_{p1} + \check{\mathscr{E}}_{k}$} \jkvv{into eq.~(\ref{eq:GammaEnsemble})} \vva{leads to} \jkvv{\ww{a \xxg{generalized finite-depth} slope-dependent}} \ww{$\Gamma_{\nabla h}$}:
\begin{eqnarray}
\hspace{-0.4cm}
\Gamma_{\nabla h} 
&=&  \jkvv{ \frac{ \langle \zeta^{2}  \rangle_{\vvk{t}} (x) }{ \mathscr{E}_{\vvx{p}1}(x) + \mathscr{E}_{\vvx{p}2}(x) + \mathscr{E}_{\vvx{k}}(x)  }\, , }
\label{eq:ee3}
\\
\nonumber
&=& \frac{ 1+   \frac{\pi^{2} \varepsilon^{2} \mathfrak{S}^{2}\Tilde{\chi}_{1}}{16} }{  1+   \frac{\pi^{2} \varepsilon^{2} \mathfrak{S}^{2} \left( \Tilde{\chi}_{1}  + \chi_{1} \right)}{32}    + \frac{\pi \nabla h }{k_{p}h_{0}}  \left[ 1 +  \xxg{\frac{\pi \nabla h  }{ k_{p}h_{0}}}  \right] \xxg{  \frac{ \yg{6\pi^{2}}\varepsilon^{2} }{ \yg{5\mathfrak{S}^{4}} (k_{p}h)^{2}}} }      \, .
\end{eqnarray}
Note that the effect of the set-down on the numerator is negligible \jfm{\footnote{because $2\langle \zeta^{2} \rangle /a^{2} \rightarrow 2 \langle ( \zeta + \langle \zeta \rangle )^{2} \rangle /a^{2}$ is corrected by an integral of $(4\langle \zeta \rangle / \mathfrak{S} h_{0})^{2} \sim \mathcal{O}(1/25\mathfrak{S}^{2})$ and the vanishing integral of $\zeta \langle \zeta \rangle$, the former being much smaller than $16 \langle \zeta \rangle/\mathfrak{S}^{2} h_{0} = \mathcal{O}(1/\mathfrak{S}^{2})$ of eq.~(\ref{eq:primsetdown}).}}.
\jkxx{Eq.~(\ref{eq:ee3}) indicates that increasing \xgg{the} slope \xgg{of} a mild \xgg{shoaling} \ww{$\nabla h < 0$} will also increase the rogue wave probability. \jkvv{Furthermore}, $\Gamma_{\vvv{\nabla h}}$ will \textit{saturate} \jkv{when \xxg{we reach} $ \nabla h _{(s)} =  - \xxg{k_{p}h_{0} / 2 \pi}$, \vvk{and} \vvv{cancel out} when $\nabla h \jkv{_{(\vv{c})}} = - \xxg{k_{p}h_{0} / \pi}$, \vvx{see \jfm{figure} \ref{fig:Ep2}.}}} \jkv{On the other hand,} \xgg{at} de-shoaling zones \ww{($\nabla h > 0)$} \xxg{following a shoal,} \xgg{eq.~(\ref{eq:ee3}) describes} \jkv{\ww{a monotonic} decrease in \jkvv{rogue wave} probability \yg{due to the piling up} when the \jkvv{down} slope is increased} \yg{because} \ww{the term} $\vvv{\check{\mathscr{E}}_{p2}}$ \ww{increases monotonically}.

\ww{\vvx{Previous} theories disregarding the slope \vvx{were} restricted} to the range $ | \nabla h | \ww{\geqslant} 1/20$. \ww{\vvv{In contrast,} the validity of our derivation is only limited by} \jkx{the assumption \jkxx{$L \vvx{| \nabla h |} / \vva{\lambda} \vvk{\lesssim} 1$} in eq.~(\ref{eq:energy2})}, \vvx{which is why} the energy \vvc{ratio} correction diverges $\jkx{|\check{\mathscr{E}}_{p2}}| \rightarrow \infty$ \ww{for} $\pi | \nabla h| /k_{p}h_{0} \gg 1 $. \vvx{Thus}, \ww{\vvx{the shoal case of} our model is valid \vvx{for}} $0 <  |\nabla h | \lesssim k_{p}h_{0} \jkv{/ \pi}$. \jkvv{\vvk{Nonetheless, }this range covers realistic conditions in the ocean, \vvk{where}} shoaling geometries with the highest slope steepness \jkx {do not exceed} $ |\nabla h| \approx 1$ \citep{Seelig1983,Piper2005}\jkvv{\vvv{. W}hile \vvk{only} $<1\%$ of slopes over ocean cross-sections \vvv{exceed} $ |\nabla h| \approx 1$\ww{, \vvv{the}
typical mean slopes \vvv{are} $<1/10$} \citep{Becker2008}.} 
\begin{figure}
    \includegraphics[scale=0.55]{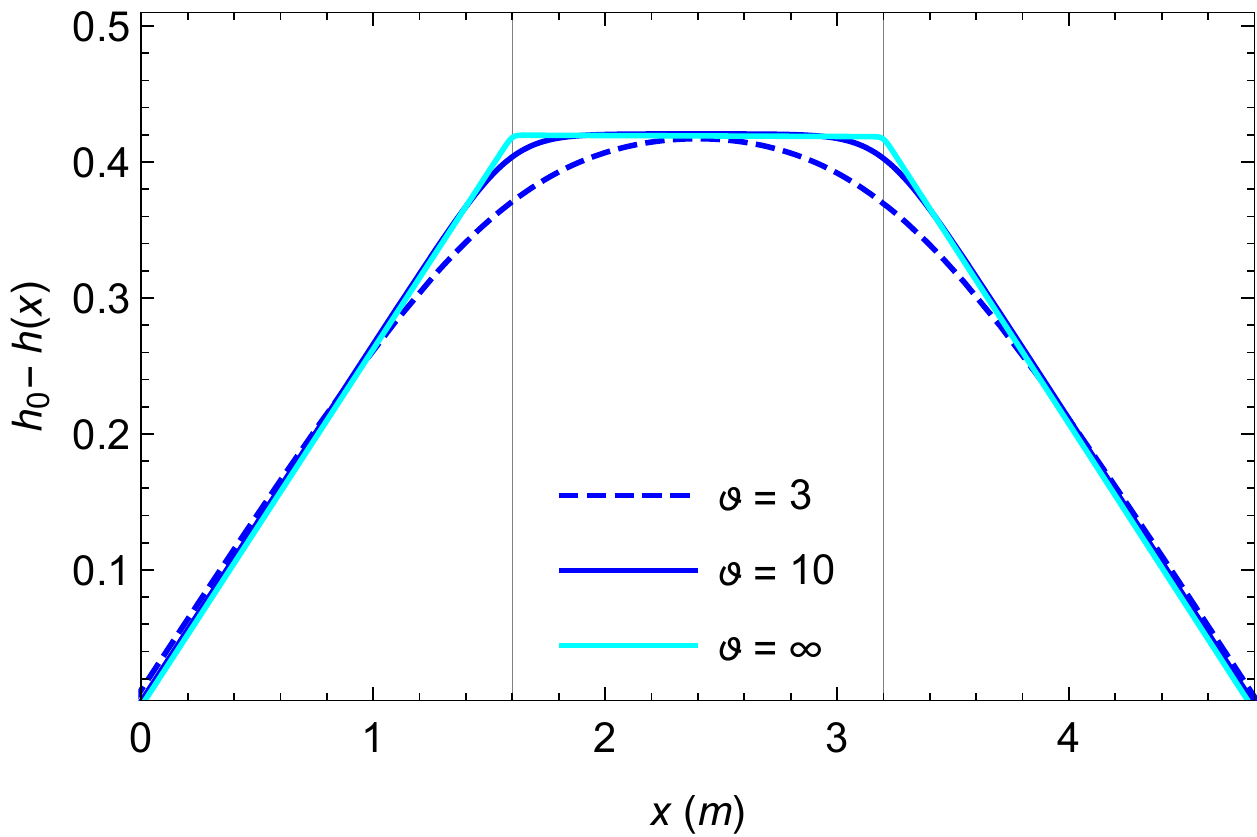}
\caption{\pp{\px{Bathymetry smoothing of the bar in \citet{Trulsen2020} generated from the integral of} the slope function $\nabla h (x)$ with finite parameter $\vartheta$ as compared to the experimental bathymetry with $\vartheta = \infty$.}}
\label{fig:smoothing}
\end{figure}

\pp{We} \jkx{compute the slope effect on $\Gamma_{\vvv{\nabla h}}$ for \jkvv{experiments} of \citet{Raustol2014} \jkvv{and} \citet{Trulsen2020}, plott\jkvv{ed} in \jfm{figure} \ref{fig:TrulsenH}.} \jkvv{For this purpose, }the evolution \vv{of} exceedance probability $\mathbb{P}_{\alpha} \vvv{\equiv} \mathbb{P} (H > \alpha  \vvx{H_{s}})$ over a shoal \jkvv{is} described as \citep{Mendes2021b}:
\begin{equation}
\jkx{  \ln  \left( \frac{\mathbb{P}_{\alpha  \, ,\,  \Gamma \ww{\,\, , \nabla h}}}{\mathbb{P}_{\alpha}} \right)  = 2\alpha^{2} \left( 1 - \frac{1}{\mathfrak{S}^{2}\Gamma_{\ww{\nabla h}}}   \right)  \, . }
\label{eq:bath}
\end{equation}
\py{A}t the peak locations \jkxx{and} \vvv{in} the de-shoaling zone\jkv{s} (\jfm{figure} \ref{fig:TrulsenH})\py{,} \vv{\py{we}} improve \pp{(cyan curve)} \py{the agreement with experimental data as compared with} the model \ww{disregarding the slope} \pp{(red curve)}\vv{, although} \jkv{t}he discrepancy in exceedance probability \jkvv{between models} displayed in all panels of \jfm{figure} \ref{fig:TrulsenH} \jkxx{does not exceed} 8\% at the location \ww{of maximum amplification of the exceedance probability}. \jkxx{This} \ww{surprisingly good fidelity of the model disregarding slope} \jkvv{is due} to \xgg{the} saturation \xgg{evidenced in eq.~(\ref{eq:ee3})}. \ww{Indeed, the experimental slope} $ \nabla h\approx - 0.26$ \jkvv{happened to be close to} \ww{saturation at} $(\nabla h)_{(s)}=- (9/5) \cdot \xxg{(1/2}\pi) \approx -0.29 $.
\pp{Although \px{our model implementing} the exact slope effect is more accurate than the steep slope approximation of \citet{Mendes2021b}, the experiments demonstrate that as long as the shoaling slope is near saturation the simpler model of \citet{Mendes2021b} is \px{already very accurate}. \px{This} provide\px{s} a physical interpretation for the success of theories based on a step \citep{Moore2020,Adcock2021c,Mendes2021b} in describing steep slope configurations.} 

\pp{However,} \vvx{w}e also checked the applicability of our model to real ocean slopes which vary \py{smoothly and} continuously, whereas the bar in the considered experiment features sharp edges. \xgg{To that purpose, }\jkvv{\ww{we \vvv{numerically smoothed the edges of the bar}} employ\vvv{ing} logistic functions \vvx{with parameter} $\vartheta$}:
\begin{eqnarray}
\hspace{-0.6cm}
\jkvv{ \frac{ \nabla h (x) }{ | \nabla h|  } =    \frac{1}{ 1 + e^{-\vartheta (x - L)  } }   \left( 1 +  \frac{1}{1+e^{-\vartheta (x - 2L)}  } \right)  - 1   \, . }
\label{eq:dh(x)}
\end{eqnarray} 
\jkv{\pp{Therefore, \px{we investigated the effect of using the smoothed slope function on} the exceedance probability \px{evolution}.} \py{W}\pp{e} verified that the \ww{amplification of the exceedance probability displayed in \jfm{figure} \ref{fig:TrulsenH} \pp{with smoothed shoal \py{edges} (dotted curve)} marginally deviates from the exact $\nabla h$} \pp{(cyan curve)} \ww{for} $\vartheta \gtrsim \vvv{3}$ \ww{corresponding to a bell-shaped bar \px{(see \jfm{figure} \ref{fig:smoothing})  }}\vvv{, and \vvx{is} indiscernible from the sharp edges when $\vartheta \xgg{\geqslant} 10$.}} \vv{This insensitivity to curvature ensures \vvx{the} applicability to real shapes in the ocean.}

\yg{F}ind\yg{ing} the excursion of the slope function \yg{in} the region of $\Gamma_{\nabla h} (\nabla h \rightarrow 0) = 1$ \vvk{would} require the analytical evolution $\varepsilon (\nabla h)$, that is unavailable \citep{Eagleson1956,Shuto1974,Walker1983,Goda1997,Srineash2018}. \vvg{Therefore, we perform a \xxg{parameterization}.} A residue $B(k_{p}h , \nabla h)$ relevant only for \vvx{very} mild slopes is introduced:
\begin{equation}
\check{\mathscr{E}}_{p2}\jkv{(k_{p}h)} \approx \frac{\pi \nabla h}{k_{p}h_{0}}  \left[ 1 + \xxg{\frac{\pi \nabla h  }{ k_{p}h_{0}}}   \right] \frac{ \yg{6\pi^{2}}\varepsilon^{2} }{ \yg{5\mathfrak{S}^{4}} (k_{p}h)^{2}} + B(k_{p}h , \nabla h)  \,\, ,
\label{eq:energycorrec44}
\end{equation}
\vvv{\vvk{noting} that} $\lim_{\nabla h \rightarrow 0} \check{\mathscr{E}}_{p2} = \lim_{\nabla h \rightarrow 0} B $ \vv{and}
\begin{equation}
\ww{\lim_{\nabla h \rightarrow 0} \Gamma_{\nabla h} = } \frac{ 1+   \frac{\pi^{2} \varepsilon^{2} \mathfrak{S}^{2}}{16}     \, \Tilde{\chi}_{1} }{  1+   \frac{\pi^{2} \varepsilon^{2} \mathfrak{S}^{2}}{32}   \, \left( \Tilde{\chi}_{1}  + \chi_{1} \right) + B(\ww{k_{p}h,}\nabla h =0)  } = 1 \, .
\label{eq:Blimit}
\end{equation}
\ww{Denoting $|\nabla h|_{-} \vv{\ll 1/20}$ and $|\nabla h|_{\jkvv{+}} \vv{> k_{p}h_{0}}$ as the \vvv{slopes} minimiz\vvv{ing} and maximiz\vvv{ing} the slope effect on the \textcolor{black}{exceedance} probability, eq.~(\ref{eq:Blimit}) imposes:}
\begin{equation}
\jkv{  B( \ww{|\nabla h|_{-}}) =   \frac{\pi^{2} \varepsilon^{2} \mathfrak{S}^{2}}{32}   \, \left( \Tilde{\chi}_{1}  - \chi_{1} \right)  > 0  \quad . }
\label{eq:B}
\end{equation}
The flat bottom boundary condition \vvv{also} requires \jfm{\footnote{\vvv{The flat bottom boundary condition requires that $\Gamma = 1$. For the most part of $\nabla h > 0$ we have a monotonically decreasing $ \Gamma_{\nabla h} <1$. However, according to eq.~(\ref{eq:ee3}), for mild slopes $\nabla h \ll 1/20$ in the de-shoaling zone we have $1 < \Gamma_{\nabla h} < \Gamma$ \vva{with} a sharp derivative $\partial \Gamma_{\nabla h}/\partial |\nabla h| > 0$ for mild slopes \vva{as} a corollary of the continuity of $\check{\mathscr{E}}_{p2}$.}}}:
\begin{equation}
\hspace{-0.2cm}
  \lim_{\nabla h \rightarrow \ww{ |\nabla h|_{-} } } \frac{\partial \Gamma_{\vvv{\nabla h}} }{\partial |\nabla h|} > 0  \,\, \therefore \,\,  \ww{\lim_{\nabla h \rightarrow \ww{|\nabla h|_{-}}} } \frac{\partial B}{\partial |\nabla h|} < 0 \,\, ,  
\label{eq:bound}
\end{equation}
\jkvv{\ww{impos\vvv{ing} the form} $B \ww{(k_{p}h,\nabla h)} =  B_{0}\vv{(k_{p}h)}|\nabla h|^{-n}$} \jfm{\footnote{\jkvv{Having $B = -B_{0}|\nabla h|^{n}$ we \vva{would} not comply with eq.~(\ref{eq:B}) because $B$ \vva{would} vanish instead of reach\vva{ing} relatively high positive values when $| \nabla h | \ll 1/10$.}}}. 
\ww{Causality and \yg{neglecting} reflection \yg{effects on $\Gamma$} \vvk{lead to}}:
\begin{figure}
    \includegraphics[scale=0.45]{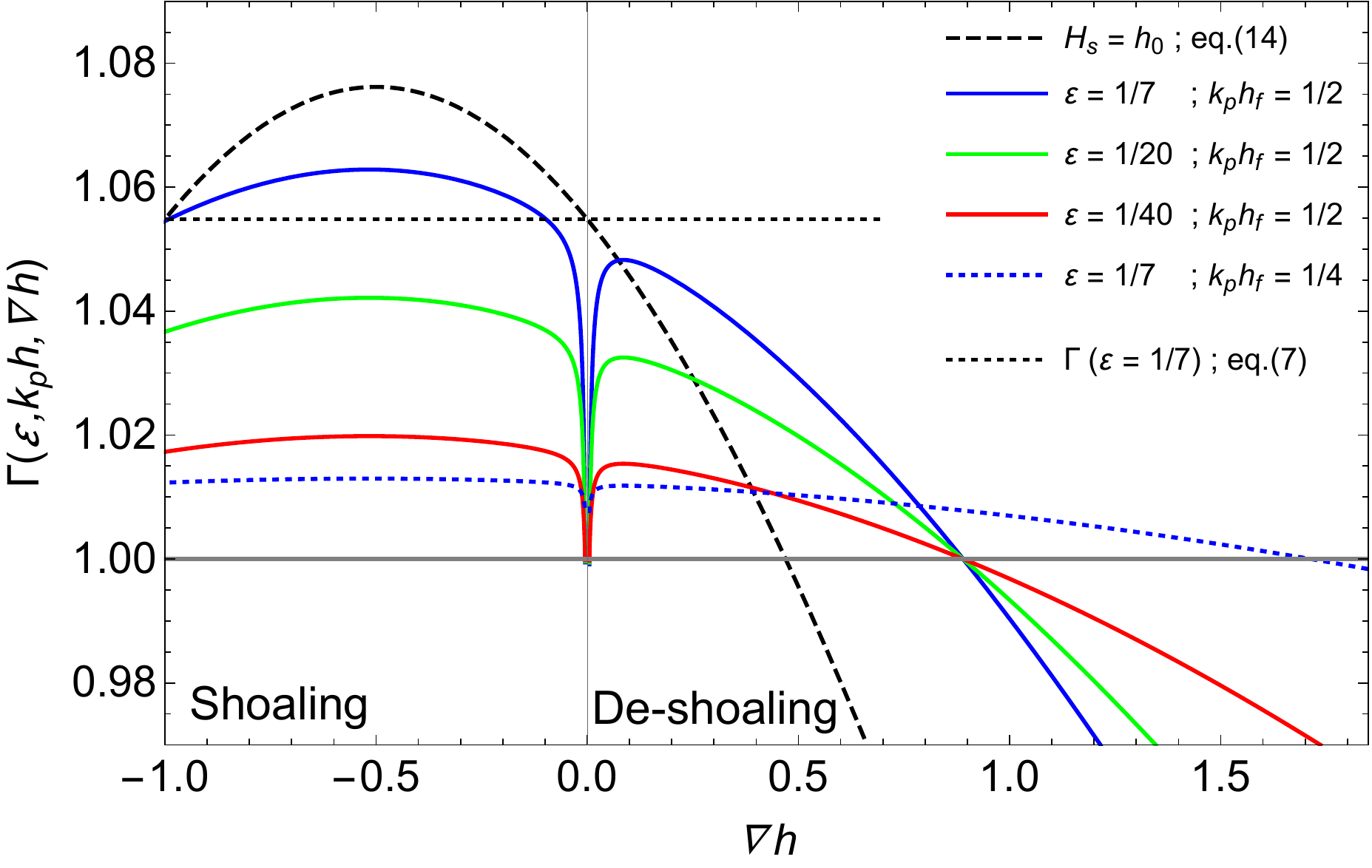}
\caption{Computation of $\Gamma_{\nabla h}$ from eq\vvx{s}.~(\ref{eq:eenablahFIM}) (solid), eq.~(\ref{eq:Ep2}) (dashed) and \citet{Mendes2021b} (dotted), \xxg{for an initial depth $k_{p}h_{0}=\pi$.}}
\label{fig:DeltaE3X}
\end{figure}
\begin{eqnarray}
\lim_{k_{p}h \rightarrow k_{p}h_{0}} \Gamma (\varepsilon , |\nabla h|_{\jkvv{+}}) = \lim_{k_{p}h \rightarrow k_{p}h_{0}} \Gamma (\varepsilon , |\nabla h|_{\jkvv{-}}) \,\, .
\label{eq:criteriaB}
\end{eqnarray}
Applying the general form of $B(k_{p}h,\nabla h)$ to eqs.~(\ref{eq:energycorrec44},\ref{eq:criteriaB}) results in:
\begin{equation}
\frac{ 6 \pi^{3} \varepsilon^{2}_{0} |\nabla h|_{+} }{ 5\mathfrak{S}^{4} (k_{p}h_{0})^{3} }  \left[ 1 \pm \frac{\pi \, |\nabla h|_{+} }{ k_{p}h_{0}}  \right]   = \pm B(k_{p}h_{0},|\nabla h|_{-})   \quad ,
\label{eq:energycorrec23}
\end{equation}
with $\pm$ denoting de-shoaling and shoaling, respectively.
To leading order in $|\nabla h|_{+} |\nabla h|_{-} \sim 10^{-2}$ we obtain:
\begin{equation}
B (k_{p}h_{0}, \nabla h) \approx  \frac{6\pi^{2}}{25\mathfrak{S}^{4}} \frac{ \pi^{2} \varepsilon^{2}_{0} }{2000 (k_{p}h_{0})^{4} }  \frac{ | \nabla h|_{-}^{n-2} }{ | \nabla h|^{n}  }   \quad . 
\label{eq:Bineq}
\end{equation}
By definition, $| \nabla h|_{-}$ corresponds to the limit in eq.~(\ref{eq:Blimit}). Having $6\pi^{2}/25\mathfrak{S}^{4}\approx 1$ \citep{Mendes2021a,Mendes2021b}, we equate eqs.~(\ref{eq:B},\ref{eq:Bineq}):
\begin{figure*}[t]
\hspace{0.5cm}
\minipage{0.45\textwidth}
    \includegraphics[scale=0.64]{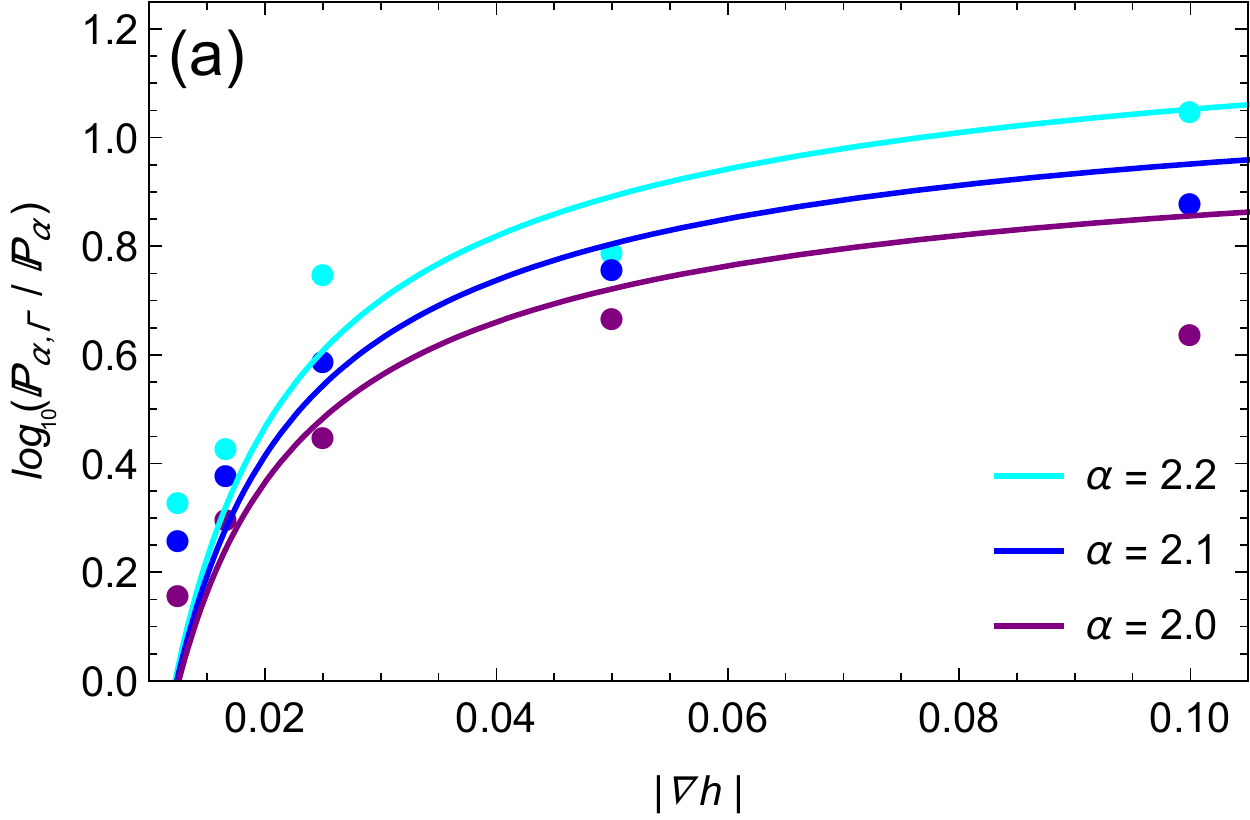}
\endminipage
\hfill
\minipage{0.5\textwidth}
    \includegraphics[scale=0.45]{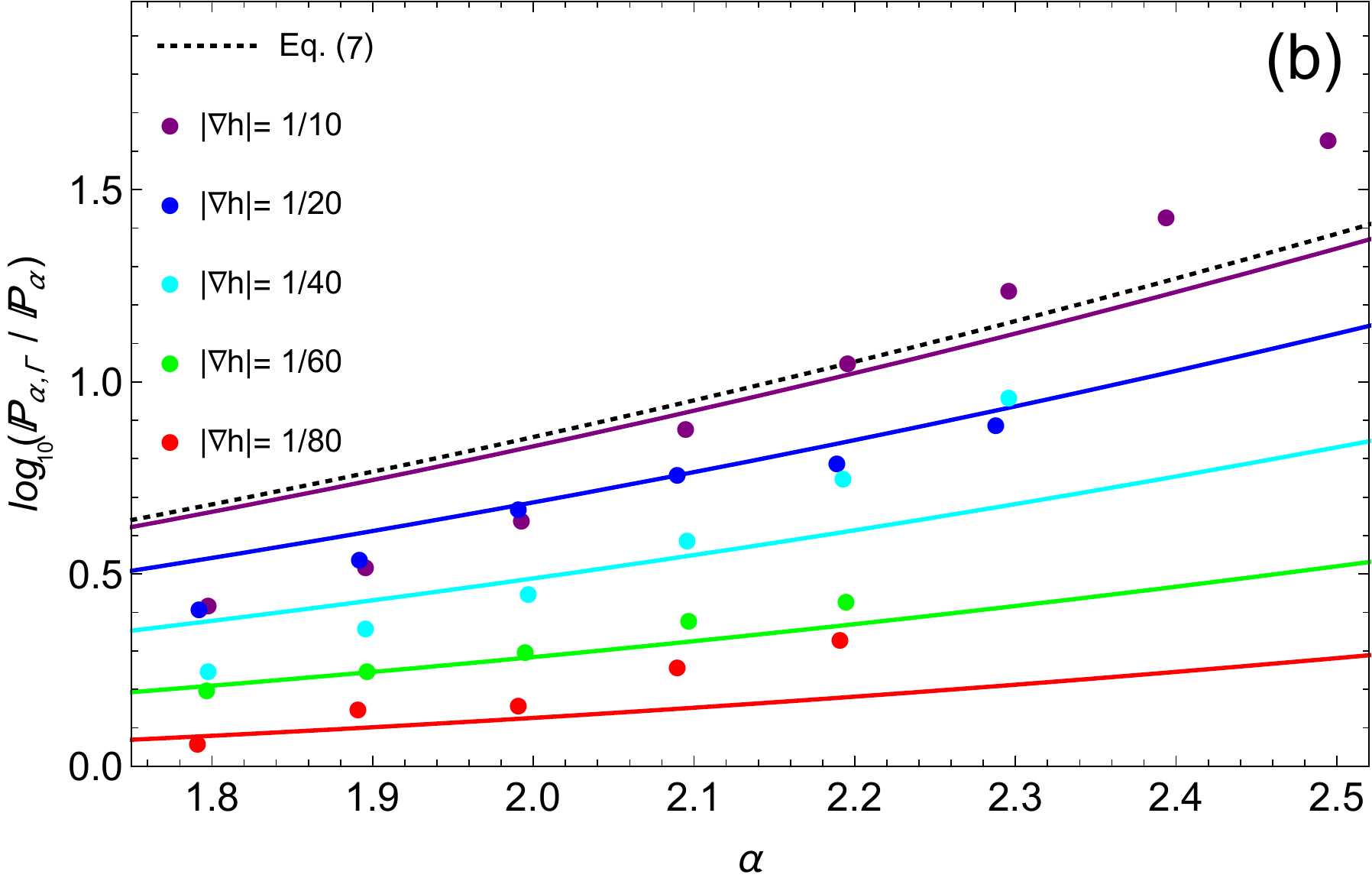}
\endminipage
\caption{\jkx{\jkv{Ratio of exceedance probabilities (relative to the Rayleigh distribution)} \jkvv{as a function of (a) \vvv{slope $\nabla h$} and} (b) normalized heights \vvv{$\alpha = H/\vvx{H_{s}}$} \jkv{reported} from \citet{Adcock2020}. \vv{D}ots \vvv{display numerical data} at $k_{p}h = 0.7$ \jkv{while} our model \jkvv{of eq.~(\ref{eq:ee3x}) \vvv{is shown in dotted line} \xxg{and} eq.~(\ref{eq:eenablahFIM})} \vvv{in solid lines.}}}
\label{fig:ZhengProb}
\end{figure*}

\begin{equation}
\vspace{-0.01cm}
  |\nabla h|_{-}   \approx \frac{\varepsilon_{0}}{\varepsilon}    \frac{ 1  }{ \sqrt{ 90 (k_{p}h_{0})^{4} (\tilde{\chi}_{1} - \chi_{1})  }  }  \quad ,
\label{eq:minSlope}
\end{equation}
so that $ | \nabla h|_{-}  \approx 1/90$ for $k_{p}h \in [0.5, 1.5]$, experimentally observed in \citet{Swan2013}. At the depth corresponding to the maximum amplification $(k_{p}h \approx 1/2)$ eq.~(\ref{eq:B}) results in $B(|\nabla h|_{-}) \approx 5\pi^{2}\varepsilon^{2}$. Since the slope effect loses importance for $|\nabla h|> 1/20$, the growth becomes $  ( \partial B / \partial |\nabla h| )_{|\nabla h|_{-}} > -120\pi^{2}\varepsilon^{2} $. Then, the derivative of eq.~(\ref{eq:Bineq}) imposes $n \approx \pi^{2}/12 \sim 1$:
\begin{eqnarray}
 B (k_{p}h, \nabla h) \approx \frac{  \pi^{2} }{25 (k_{p}h_{0})^{2} }  \frac{\varepsilon^{2}}{(k_{p}h)^{2} | \nabla h|}\,\, . 
\label{eq:Bineq3}
\end{eqnarray}
The intermediate depth energy ratio reads:
\begin{equation}
\check{\mathscr{E}}_{p2} \approx \frac{5\varepsilon^{2}}{(k_{p}h)^{2}} \left\{ \frac{ \pi \nabla h}{k_{p}h_{0}}  \left[ 1 + \frac{\pi \nabla h  }{ k_{p}h_{0}}   \right] +  \frac{ \pi^{2} }{125(k_{p}h_{0})^{2} | \nabla h | }   \right\} .
\label{eq:energycorrec22XX}
\end{equation}
\vvv{Plugging eq.~(\ref{eq:energycorrec22XX}) into eq.~(\ref{eq:ee3}) results in (\jfm{figure} \ref{fig:DeltaE3X}),}
\begin{widetext}
\begin{equation}
\vvx{  \Gamma (k_{p}h,\varepsilon,\nabla h) =  \frac{ 1+   \frac{\pi^{2} \varepsilon^{2} \mathfrak{S}^{2}}{16}     \, \Tilde{\chi}_{1} }{  1+   \frac{\pi^{2} \varepsilon^{2} \mathfrak{S}^{2}}{32}   \, \left( \Tilde{\chi}_{1}  + \chi_{1} \right) + \frac{5\varepsilon^{2}}{(k_{p}h)^{2}} \left\{ \frac{ \pi \nabla h}{k_{p}h_{0}}  \left[ 1 + \xxg{\frac{\pi \nabla h  }{ k_{p}h_{0}}}   \right] +  \frac{ \pi^{2} }{ \xxg{125}(k_{p}h_{0})^{2} | \nabla h | }  \right\} }  \quad . }
\label{eq:eenablahFIM}
\end{equation}
\end{widetext}
\vvv{The cancelling effect of $B(k_{p}h, \nabla h)$ \xxg{on} the \vvc{pre-shoal} flat bottom $\Gamma$ clearly appears \xgg{around $\nabla h = 0$ in} \jfm{figure} \ref{fig:DeltaE3X}.} \jkx{However,} \jkvv{when} $k_{p}h \rightarrow 0$ the trigonometric coefficients $(\chi_{1} , \tilde{\chi}_{1}) \rightarrow  \infty$ \jkxx{grow much faster than $\jkxx{\check{\mathscr{E}}_{p2}}$}, \jkvv{and $\Gamma$} no longer depend\jkvv{s} on \jkvv{$\nabla h$}. \jkx{\jkv{This \xgg{slow} dependence on the slope \jkvv{in} shallow waters (\vvk{see the \xgg{blue} dotted curve} in \jfm{figure} \ref{fig:DeltaE3X}) has been} observed in \citet{Doleman2021}\xxg{, which carrying experiments at $k_{p}h_{0} \approx \xgg{0.38}$ found the evolution of the kurtosis for a step ($|\nabla h |= \infty$) \xgg{and} a slope of $|\nabla h |= \xgg{0.05}$ to be identical. Our model explains this phenomenon \xxg{with} the proximity of the slope to the saturation point $ | \nabla h |_{(s)} =   k_{p}h_{0} /\xxg{2}\pi \approx \xgg{0.06}$.}} 

\pp{Experiments with wide ranges of slopes, \py{i.e. with length} $L \sim \py{\lambda}$ and  water depth $\py{\pi/10 \lesssim } \, k_{p}h \, \py{\lesssim \pi/2}$, are \px{not available to date} because mild slopes usually require lengths exceeding wave tank dimensions or wave frequencies must be too high for the given dimensions. Hence, in the absence of experiments with broad ranges for slopes, steepness and water depth,} we \ww{assess} our theory \ww{against} \vvv{the} numerical results \jkv{of} \citet{Adcock2020}, describ\jkv{ing} how the probability of the envelope \jfm{\footnote{\jkv{The envelope is computed as $R = \sqrt{\zeta^{2} + \hat{\zeta}^{2}}$, with $\hat{\zeta}(x,t)$ the Hilbert transform of the surface elevation $\zeta(x,t)$. We may approximate $R \approx H \approx 2a$ \xgg{which} leads to $a/\sigma \approx 2\alpha = 2H/\vvx{H_{s}} $, where $\sigma^{2} = \int_{0}^{\infty} S(\omega) \, d\omega $.}}} is affected by increasing the slope steepness. In figure 13 of \citet{Adcock2020} \ww{the shoal increased the} exceedance probability \ww{for rogue waves as soon as} $| \nabla h | = 1/80$, with a saturation of this effect for slopes steeper than $| \nabla h | \gtrsim 1/10$ (the details of physical variables of the performed simulations C1-C8 are found on table II in \citet{Adcock2020}).  \vvv{W}e \jkv{apply the \ww{same} conditions to} the slope-dependent non-homogeneous correction \jkv{\vvv{of} eqs.~(\ref{eq:ee3x}), (\ref{eq:ee3}), (\ref{eq:eenablahFIM}) and (\ref{eq:bath})} \vvv{and compare the maximum amplification} $(\varepsilon = 1/16)$ \vvv{of the exceedance probability (\jfm{figure} \ref{fig:ZhengProb}).} \ww{Our model reproduces well the exceedance probability for rogue waves $(\alpha > 2)$ and its saturation \vvv{for steep slopes} (\jfm{figure} \ref{fig:ZhengProb}\jfm{a}) \xgg{or for large waves $\alpha \geqslant 1.75$ with fixed slope} \vvx{as shown} in \jfm{figure} \ref{fig:ZhengProb}\jfm{b}}. Furthermore, \vv{we recover} our \jkv{previous} model \citep{Mendes2021b} for the steepest slopes \ww{(see dotted line in \jfm{figure} \ref{fig:ZhengProb}\jfm{b})}. 

\section{DISCUSSION}

\uu{The slope effect on the exceedance probability can be interpreted as a second redistribution o\pp{f} the wave statistics\px{, on top of that induced by the depth change}. In the presence of a strong departure \pp{from} a zero-mean water level due to a set-down/set-up \pp{\px{the} potential energy density \px{is} affected by \px{a} slope\px{-induced} correction $\check{\mathscr{E}}_{p2}$}. In the case of a shoal, \pp{such energy disturbance} decreases the total potential energy as compared to linear homogeneous waves, thereby increasing the effect of the energy redistribution \pp{($\Gamma_{\nabla h} > \Gamma$)}. Similarly, a set-up \pp{induced by} wave-breaking would cause the total potential energy to increase, and so we would observe the opposite effect by decreasing the exceedance probability \pp{because of $\Gamma_{\nabla h} < \Gamma$}.} \px{This means that the depth change has the leading order in amplifying the exceedance probability over a shoal when the slope steepness does not vanish ($|\nabla h| \rightarrow 0$), while \py{the slope \pz{modulates}} the energy redistribution due to th\py{is} depth change.}

\uu{The saturation \pp{of the slope} effect can be understood as a combination of the effect of lowering the mean water level as a function of the slope and \px{of} the pace of the depth transition itself. The secondary term in brackets of eq.~(\ref{eq:Ep2}) is equivalent to $\langle h(x) / h_{0} \rangle$\py{. A} \pp{continuous} steep slope \px{$|\nabla h| \rightarrow \infty$} \pp{implies} $\langle h(x) / h_{0} \rangle \rightarrow 0$ \pp{\px{over the wave relaxation region following the start of the shoal. Indeed, over this region} the mean depth converges to the shallower depth}. In the meantime, a very steep slope will quickly increase the set-down. \pp{Nevertheless,} the fast increase \pp{in the set-down} is balanced by the fast decrease in \pp{mean depth}, therefore creating \px{the observed} saturation of \pp{their product, namely} $\check{\mathscr{E}}_{p2}$. \pp{In other words, the response of the set-down to the steep slope transition past the saturation point is slower than the depth transition itself and has no time to develop. \px{Conversely, i}n the de-shoaling zone} the faster increase \pp{of the set-up} due to steeper slopes is not balanced by the depth \pp{transition}, as \pp{the mean depth} will increase \py{rather than decrease.}}

\uu{Our framework poses a clear unifying picture for wave statistics and energetics transitioning from deep to shallow waters: (i) waves in deep water will not have \pp{their} energy affected by the slope and tend to follow Gaussian statistics, (ii) in intermediate water the wave energy density will be redistributed due to depth effects on the steepness\py{,} vertical asymmetry and mean water level\py{, ultimately} increas\py{ing} rogue wave likelihoods\px{,} (iii) in shallow water the effects on steepness and vertical asymmetry still exist, but the quick divergence of \pp{the superharmonics} halts the energy \xxk{redistribution} \pp{while} the set-up inverts \pp{the latter}. \pp{Therefore,} in the absence of any ocean process besides shoaling, we unify within a single physical mechanism the \pp{seemingly} contradictory results of \citet{Higgins1952} in deep water, \citet{Trulsen2020} in intermediate water and \citet{Glukhovskii1966} in shallow water.}

\section{Conclusions}

We have for the first time \vv{obtained} an analytical dependence \vvv{of the \vvk{wave height} exceedance probability} on $\nabla h$. \uu{It widely extends the approach developed for steps and unifies the theories for wave statistics in deep, intermediate and shallow waters within a dynamical evolution. The unified framework is laid out as bathymetry effects on the energy partition between waves of different heights, and therefore the probability distribution, by considering the specific effects of the slope beyond the sole bathymetry change.} \jkv{\vvx{Models that do note take finite slopes into account are nonetheless capable of reproducing well the wave statistics of steep slopes} \citep{Adcock2021c,Moore2020,Mendes2021b}.} \vvk{ We explain \vvc{this}} equivalency \vv{between a step and steep slopes} \vvx{\vvc{with} the saturation effect \xgg{as evidenced from eq.~(\ref{eq:ee3})}}. \yg{When slopes become too steep and we reach saturation,} \vvx{the \yg{success of these models} can be interpreted as the result of the slope effect being \yg{fu}lly encoded in} the change \vvx{of} \yg{both steepness and depth}. Although \vv{\ww{our} model} does not cover the limiting case of a step \textit{per se}, both \xgg{steep} and mild bathymetric profiles in the ocean are well covered by the model range of validity. \vvg{Furthermore, our range of validity is consistent with small reflection effect due to a non-diverging surf similarity.} \jkxx{Qualitatively, our theory points to three major consequences. Firstly, making a mild} slope steeper \jkxx{increase\vvx{s} the} probability of \vv{large} wave heights in shoaling zone\vvv{s}, and decrease\vvv{s it} in de-shoaling zone\vvv{s} \xxg{following a shoal}. 
\yg{Secondly}, in very shallow water the slope effect \vv{already} saturates even for mild slopes, while in \vv{intermediate} waters the saturation point is \jkv{harder to attain}. Thirdly, \yg{we reconcile the transient increase of rogue wave probability over a shoal with lower probabilities in shallow water.}
\jkxx{\vvv{W}e have quantitatively \ww{validated our model against} the numerical results of \citet{Adcock2020} and the experiments of \citet{Raustol2014} for the exceedance probability of wave heights, \ww{obtaining} good agreement. \pp{Finally, \px{the} unification \px{of} rogue wave formation mechanisms within the present framework \px{should be} possible, provided future work addresses the out-of-equilibrium ocean processes driving non-Gaussian statistics over a flat bottom, such as opposing currents and crossing seas.}}

\section{Acknowledgments}

S.M\vvc{.} and J.K. were supported by the Swiss National Science Foundation under grant 200020-175697. \vvg{We thank Maura Brunetti for fruitful discussions.}

\bibliography{Maintext}

\end{document}